# Vibrational sum frequency scattering in absorptive media: A theoretical case study of nano-objects water

S. Kulik, S. Pullanchery, and S. Roke[,*]

Laboratory for fundamental BioPhotonics, Institutes of Bioengineering (IBI) and Materials Science (IMX) and Engineering, School of Engineering (STI), and Lausanne Centre for Ultrafast Science, École Polytechnique Fédérale de Lausanne (EPFL), CH-1015 Lausanne, Switzerland.

[*]Corresponding author. E-mail: sylvie.roke@epfl.ch

**ABSTRACT**

The structures of interfaces of nano- and microscale objects in aqueous solution are important for a wide variety of physical, chemical and biological processes. Vibrational sum frequency scattering has emerged as a useful and unique probe of the interfacial structure of nano- and microscale objects in water. However, the full surface vibrational stretch mode spectrum has not been measured yet, even though it would be extremely informative to do so. The reason for this is that probing the vibrational modes of interfacial water requires a full understanding of how the linear absorptive properties of the bulk aqueous medium influence the sum frequency scattering process. Here, we have simulated vibrational sum frequency scattering spectra of the interface of nanoscale objects dispersed in water. We analyzed the effect of the infrared pulse absorption on the outcome of surface vibrational sum frequency scattering measurements. We find that both infrared absorption as well as the type of optical detection can drastically modify the measured vibrational interfacial spectrum. The observed changes comprise spectral distortion, frequency shifting of the main vibrational stretch mode and the introducing of a new high frequency peak. This last feature is enhanced by non-resonant interactions.

**Introduction**

Aqueous interfaces are abundant in nature and important for a host of processes related to atmospheric[1-2], chemical[3], biological[4] and geological[5] transformations. The molecular and structural properties of aqueous interfaces are therefore important. Measuring these is challenging, as one has to separate the probed properties of the small minority of interfacial water molecules, from the very large amount of bulk water in a typical sample. The demonstration of surface vibrational sum frequency generation (SFG) spectroscopy[6-7] therefore represented a great improvement as it allowed the selective spectroscopic identification of interfacial water in a background-free way. Since then, vibrational SFG has been used in reflection geometry to probe planar extended interfaces of water and salt solutions in contact with air[8-14], lipids[15-20], liquids[21-24], polymers[25-28], proteins[29], quartz and silica[30-33], metals[34-36], and many others (see e.g. further reviews in Refs.[37-44]).

The next step in understanding the complexity of aqueous interfaces is to probe realistic nano- and microscale sized interfaces rather than a planar extended model interface, as shape, size and environment influence the molecular and macroscopic properties of a system. To do so, vibrational sum frequency scattering (SFS)[45-46] was invented. Instead of reflecting infrared (IR) and visible (VIS) laser pulses from a planar extended interface, both beams are passed through a sample (Fig. 1a), typically a liquid that contains particles, droplets or liposomes, and the interfacially generated sum frequency (SF) photons are scattered and detected. In this way the interfaces of small objects in the size range ~ 20 nm – 50 µm can be probed on the molecular level. While the surface vibrational spectrum provides information on the molecular species and the local environment, the scattering pattern as detected in different polarization combinations provides additional information on molecular orientation[47], surface electrostatics[48-49], aggregation[50] and chirality[51]. Vibrational SFS spectra have been recorded in many spectral ranges, starting with probing C-H modes (from stearyl-coated silica particles in $CCl_4$ or other hydrophobic liquids[52-53]) as this is typically the sweet spot of the used ultrafast infrared laser source. Spectra were soon recorded of diverse objects in aqueous solution addressing a wider spectral range that contains vibrational C-H stretching[54-56], bending[57], amide bands[57], P-O stretching[58], -CN stretching[59], S-O stretching[54] and other modes. The surface vibrational spectrum of water (using $D_2O$) was measured by SFS in the form of water droplets dispersed in a hydrophobic liquid[60]. The inverse, the measurement of the vibrational water spectrum of an object dispersed in aqueous solution, has not been measured: Water spectra have only been measured up to the edges of the high frequency side of the absorption band in $D_2O$[61] (from 2650 - 2750 cm-

[1]) and in a $H_2O:D_2O$ mixture[62] (from 2600 – 2750 $cm^{-1}$), excluding the biggest part of the vibrational stretch mode region of water (2200 – 2750 $cm^{-1}$ for D-O stretch modes).

Thus, even though a vibrational water spectrum of a nanoscale interface embedded in water would contain a wealth of information about interfacial water structure and hydrogen bonding interactions, it has not been measured. The reason for this lack of information is that water is a highly absorptive medium for infrared pulses. Therefore it is expected to be very challenging to measure and interpret a sum frequency scattering spectrum of particles, liposomes, or droplets dispersed in water. Fig. 1B illustrates what happens to an IR pulse as it travels through water: After a distance on the order of a micrometer the IR beam will be attenuated significantly as the IR light is absorbed by the water. This attenuation is frequency dependent, distorting the spectral contents of the IR light. Fig. 1C shows IR absorption spectra of 1:1 mixtures of $H_2O$ and $D_2O$ for several different optical path lengths ranging from 1 µm to 50 µm. Taking D-O stretch modes as the vibrational modes of interest, for the strongest absorption at a frequency of ~ 2500 $cm^{-1}$ after 10 µm penetration all intensity is absorbed, while for ~2750 $cm^{-1}$ there is still > 50 % of the original intensity after 50 µm of path length. This means that, as an IR pulse travels through an absorptive medium such as water its intensity attenuates drastically and its spectral shape is continuously modified. In order to understand vibrational SFS spectroscopy in an absorptive medium a computational analysis is desirable.

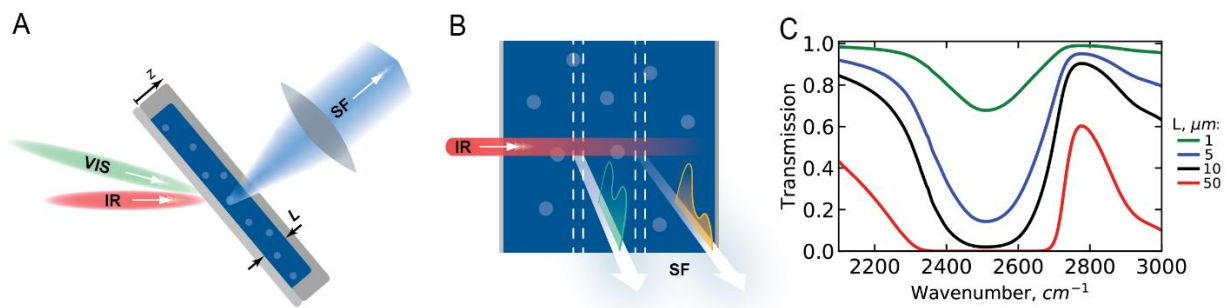

**Figure 1: Linear optical interactions distort SFS spectra** (A) Sketch of the SFS experiment. The direction of the outgoing SF light is perpendicular to the cuvette, and indicated as z. L is the thickness of the cuvette. (B) Illustration of linear IR absorption in an SFS experiment aimed to measure interfacial water in an aqueous medium. (C) The IR transmission profiles for different slab thicknesses of a 1:1 $D_2O:H_2O$ mixture.

Here, we describe a computational study of sum frequency scattering of interfacial O-D vibrational stretch modes of a scattering object in an absorptive medium. The presented theory is valid when the probed surface is composed of the same chemicals as the bulk medium. We focus on the example of sub-micron sized / nanoscale scattering objects (e.g. particles or droplets or bubbles or liposomes) in an aqueous medium, and calculate the vibrational SFS response of the interfacial water. We assume single scattering, valid for the experiments that have so far been published in the literature. We compute scattered sum frequency spectra of the vibrational modes of interfacial water, given a certain defined surface response and determine the modulation of the surface optical response by absorption of the IR pulse. We then consider the effect of the presence of a non-resonant interfacial response and the size of the scattering volume, determined by optics that collects the generated sum frequency photons and the linear scattering in the turbid medium. We find that IR absorption, a non-resonant background and the choice of the scattering volume from which the SF light is detected can all drastically modify the spectral shape of the scattered SF light.

**Results and Discussion**

**Approach.** In order to perform the computational analysis, we start with the expression for the SFS intensity that is emitted from a certain scattering volume[63]. For a sample with an optical path length $L$ for the SF light, in the $z$-direction (Fig. 1A), the intensity of scattered SF photons ($I_{SF}(\omega_{SF}; \omega_{VIS}, \omega_{IR})$) is given by:

$$I_{SF}(\omega_{SF}; \omega_{VIS}, \omega_{IR}) \propto \int_0^L |\mathbf{\Gamma}^{(2)}(\omega_{SF}; \omega_{VIS}, \omega_{IR})|^2 I_{IR}(\omega_{IR}, z) I_{VIS}(\omega_{VIS}, z) f_{focal}(z) \rho(z) \, dz, \qquad (1)$$

where $\omega_{SF}, \omega_{VIS}$, and $\omega_{IR}$ are the frequencies of the SF, VIS, and IR pulses, $I_{IR}(\omega_{IR}, z)$ is the intensity of the incident infrared illumination, $I_{VIS}(\omega_{VIS}, z)$ is the intensity of the incident visible illumination, $f_{focal}(z)$ is a collection function that describes the efficiency of light collection along the optical axis $z$ of the collection optics, and $\rho(z)$ is the number density of particles at a specific depth. $\mathbf{\Gamma}^{(2)}(\omega_{SF}; \omega_{VIS}, \omega_{IR})$ is the effective second-order particle susceptibility[64] that describes the spectral interfacial response of the object dispersed in solution. $\mathbf{\Gamma}^{(2)}(\omega_{SF}; \omega_{VIS}, \omega_{IR})$ is a function of the scattering angle (θ, defined as the angle between the scattered SF wavevector and that of the phase-matched direction), the size of the object (typically indicated by R, the radius), and the second-order surface susceptibility ($\chi^{(2)}$). Note that Eq. (1) is written in terms of second-order effects, but can be expanded to include effective third-order processes[48-49]. Since the source of the interfacial response is not the topic of the present study, we will use a spectral function for $|\mathbf{\Gamma}^{(2)}(\omega_{SF}; \omega_{VIS}, \omega_{IR})|^2 = |\mathbf{\Gamma}^{(2)}|^2$ as input for Eq. (1). This means that the values of R, θ and $\chi^{(2)}$ are not relevant here. Eq. (1) can be solved by numerically computing the function under the integral for small $dz$ distances (taken here as 1 µm), which are then summed. The optical distance $L$ is taken to be 100 µm to resemble a typical sample cell used for SFS measurements[63]. The bulk medium is (somewhat arbitrarily) chosen to be a 1:1 mixture of light and heavy water as it is less absorptive then pure $H_2O$ or $D_2O$.

For the description of the surface response we take an input function for $|\Gamma^{(2)}|^2$ which is a spectral response that is based on reported responses for planar aqueous interfaces in the literature[65], since vibrational surface SF spectra of small objects dispersed in water have not yet been measured. In an actual experiment, $\Gamma^{(2)}$ is the physical quantity that one wants to retrieve. Here, we take a certain surface response (plotted in Fig. 2A that resembles a typical planar water SF surface spectrum in a 1:1 mixture of light and heavy water[65]) and examine how the spectral shape of this response is modified by the linear optical interactions in the experiment. We model both C-H and O-D stretch modes and for convenience represent $|\Gamma^{(2)}|^2$ by Gaussians (O-D modes) and Lorentzians (C-H modes) centered in the O-D and C-H stretch regions, resembling the SF spectral response from a typical aqueous interface. Note that, since we are interested in the difference between the input spectral function and the output SF light the actual choice of the form of $|\Gamma^{(2)}|^2$ is not very important. $|\Gamma^{(2)}|^2$ of the O-D modes (blue curve, Fig. 2A) is modelled by a broad Gaussian centered at 2450 cm$^{-1}$ with a full width at half maximum (FWHM) of 292 cm$^{-1}$, which resembles published spectral data[60, 66]. $|\Gamma^{(2)}|^2$ of the C-H modes (green curve, Fig. 2A) is represented by 4 Lorentzians centered at 2850 cm$^{-1}$ (with a FWHM of 20 cm$^{-1}$, and relative amplitude of 0.3), 2878 cm$^{-1}$ (with a FWHM of 30 cm$^{-1}$, and relative amplitude of 0.4), 2939 cm$^{-1}$ (with a FWHM of 30 cm$^{-1}$, and relative amplitude of 0.5) and 2959 (with a FWHM of 30 cm$^{-1}$, and relative amplitude of 0.5). These modes represent the symmetric methylene stretch mode, the symmetric methyl stretch mode, the Fermi-resonance and the asymmetric methyl stretch mode, respectively. The effect of a non-resonant background is considered by adding a constant to the spectral response.

For measurements in a non-absorptive medium, the observed SF signal is normalized by the intensities of IR and visible beams incident on the sample, as has been described earlier[55]. In this case, Eq. (1) reverts to the familiar expression: $I_{SF}(\omega_{SF}; \omega_{VIS}, \omega_{IR}) \propto |\Gamma^{(2)}(\omega_{SF}; \omega_{VIS}, \omega_{IR})|^2 I_{IR}(\omega_{IR}) I_{VIS}(\omega_{VIS})$. For the visible beam ($I_{vis}$) there is an attenuation effect caused by linear light scattering which was described earlier for second harmonic scattering[67], and SFS[63]. For this we would have $I_{vis}(\omega_{vis}, z) = I_0(\omega_{vis})e^{-\tau_{vis}z}$ with $\tau_{vis}$ the turbidity. The same applies for the IR and SF beams. However, in what follows we consider a small refractive index difference between the particles and the bulk medium (known as the Rayleigh-Gans-Debye approximation[51]), which means we neglect attenuation through linear scattering. This is appropriate since the effect of IR absorption is much bigger. For the measurement of an interfacial water spectrum of particles dispersed in water, approximating $I_{IR}(\omega_{IR}, z)$ as the infrared pulse that is generated by the light source[63] is insufficient, as the IR spectral profile attenuates and changes drastically when the beam propagates through the sample (illustrated in Fig. 1B). Therefore, the measured SF intensity cannot simply be normalized by the incident IR beam profile or an IR beam profile multiplied with the IR transmission spectrum. Instead we have to take:

$$I_{IR}(\omega_{IR}, z) = I_0(\omega_{IR})e^{-\alpha(\omega_{IR})z} \qquad (2)$$

where $I_{IR}(\omega, z)$ is the IR intensity at z, $I_0(\omega_{IR})$ is the incident intensity and $\alpha(\omega_{IR})$ is the absorption spectrum. $\alpha(\omega_{IR})$ is computed from the measured IR transmission spectrum (T, Fig. 1C black line) using Lambert-Beer's law, $\alpha(\omega_{IR}) = \frac{-\ln(T(\omega_{IR}))}{L}$. $T(\omega_{IR}, L = 10 \ \mu m) = I_{IR}(\omega_{IR}, L = 10 \ \mu m)/I_0(\omega_{IR}))$ of a 10 μm-layer of a 1:1 D$_2$O:H$_2$O mixture with a spectrally flat $I_0(\omega_{IR})$. In the experiments $I_0(\omega_{IR})$ is not spectrally flat but composed of femtosecond broadband laser pulses. To model these, for probing C-H modes we use a 200 cm$^{-1}$ FWHM Gaussian, centered at 2900 cm$^{-1}$ for $I_0(\omega_{IR})$. For the O-D modes we use 4 Gaussians with 200 cm$^{-1}$ FWHM centered at 2600, 2500, 2400, 2300 added together for $I_0(\omega_{IR})$ to simulate an SFS experiment in which the bandwidth of a single IR pulse envelope is smaller than that of the OD vibrational stretch mode band as is the case in most experiments[60]. Both spectra are plotted in Fig. 2A as the red dashed-dotted line and the red dotted line, respectively.

      Eq. (1) also contains the particle density as a function of penetration depth, which we consider as constant ($\rho(z) = \rho$), and the scattering volume, determined by the collecting lens focal depth function ($f_{focal}(z)$). We will initially assume that $f_{focal}(z) = 1$ for $0 < z < L$, and 0 elsewhere, meaning that we assume that the collection lens does not influence the experiment and the scattered SF light that is generated at every point in the sample is detected. We will first consider the effect of absorption on the SFS result, and then examine the effect of adding a non-resonant contribution. Finally, we will consider the effect of using a very small collection focal volume such as from an objective, or a larger one, such as from a small focal volume singlet lens.

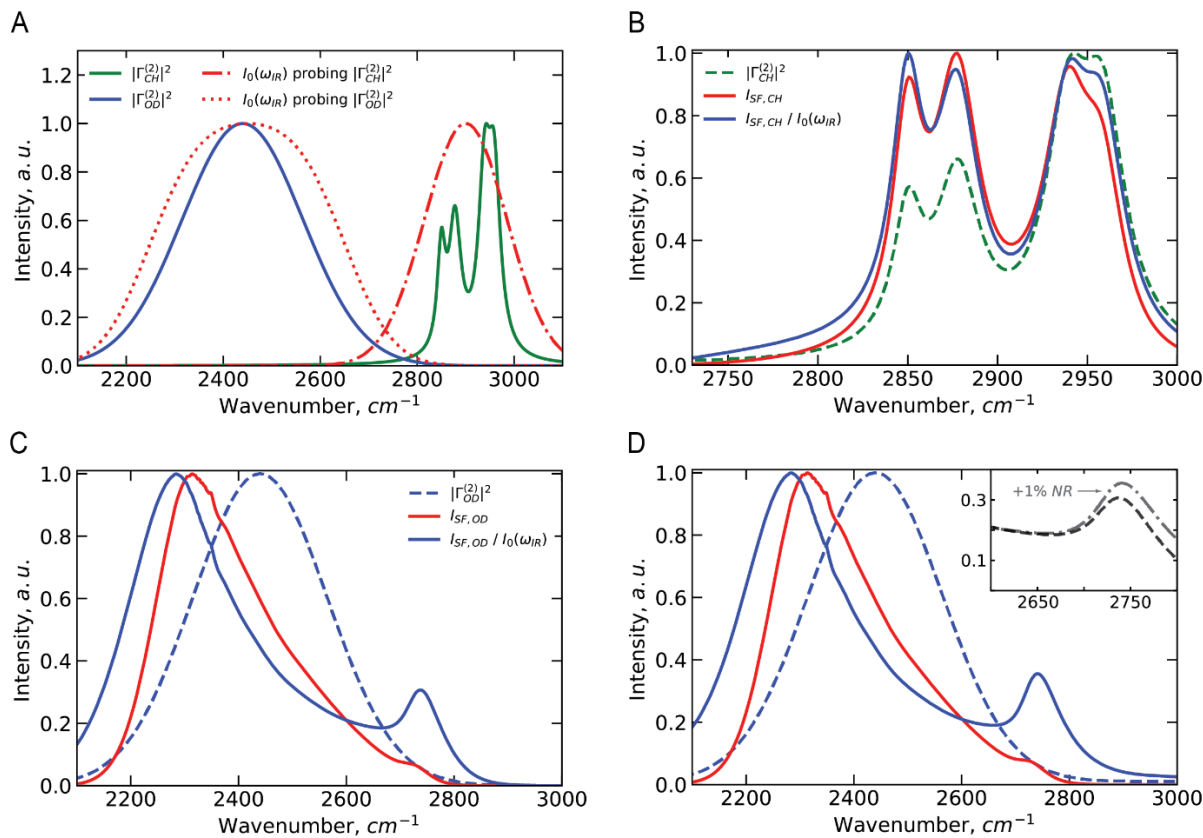

**Figure 2: Effect of IR absorption** (A) Inputs used for $\left|\boldsymbol{\Gamma}^{(2)}\right|^2$ and $I_0(\omega_{IR})$ in Eq. (1). The red lines represent the spectral shapes of the IR pulses ($I_0(\omega_{IR})$) used to probe the O-D (dotted line) and C-H (dash-dotted line) stretch mode regions. The green curve represents $\left|\boldsymbol{\Gamma}^{(2)}\right|^2$ for the C-H modes and the blue curve represents $\left|\boldsymbol{\Gamma}^{(2)}\right|^2$ for the O-D modes. (B) Effect of IR absorption on the model C-H band. The simulated signal (solid red line) is the resulting SF intensity that would be measured, and the solid blue line is the SF intensity divided by the spectral profile of the IR pulse. Although the ratio of the asymmetric and symmetric modes changes, the ratio of the symmetric $CH_3$ and $CH_2$ modes, that is usually used to determine alkyl chain order, is not changing much. (C) Effect of IR absorption on the model O-D band. The simulated signal (solid red line) is the resulting SF intensity that would be measured, and the solid blue line is the SF intensity divided by the spectral profile of the IR pulse. (D) Effect of IR absorption on the model O-D band, including a non-resonant background. The non-resonant response is incorporated by adding a constant that has a magnitude of 1 % of the maximum of the blue spectrum (blue dashed line). The simulated signal (solid red line) is the resulting SF intensity that would be measured, and the solid blue line is the SF intensity divided by the spectral profile of the IR pulse. The inset shows the amplifying influence of a non-resonant background on the high frequency peak-like feature.

**The effect of IR absorption.** Having described our approach, we consider the results of the computation. First, we discuss the influence of absorption of the IR pulse. Figure 2B shows the input function for $\left|\boldsymbol{\Gamma}^{(2)}\right|^2$ (green curve) for the C-H modes. Computing Eq. (1) for this spectral region results in the simulated signal represented by the red line ($I_{SF,CH}$). This includes both multiplying by the spectral shape of the incoming pulses, $I_0(\omega_{IR})$, as well as correcting for IR absorption (Eq. (2)). Both effects combined result in a spectral difference compared to the input $|\Gamma^{(2)}|^2$. It is common in the SFS literature[55] to report the measured intensity divided by the IR pulse ($I_{SF,CH}/I_0(\omega_{IR})$). This curve is shown as the blue line in Fig. 2B. It can be seen that the additional effect of IR absorption by the H$_2$O/D$_2$O mixture changes the relative intensities for the C-H modes, even though the amount of IR absorption > 2800 cm$^{-1}$ is relatively low compared to that in the region 2200 – 2800 cm$^{-1}$. The amplitude ratios of the symmetric CH$_2$ and CH$_3$ stretch modes (which report on alkyl chain order[68-69]) as well as the amplitude ratios of the CH$_3$ symmetric and asymmetric stretch modes (which report on the alkyl chain tilt angle[70]) are influenced by it. The difference in amplitude is 9 % and 20 % respectively, which would amount to an approximate change in tilt angle of < 7°. Note that this estimation was made taking ratios of a recent study[58] and ignoring the fact that this study was conducted in pure D$_2$O. Since most C-H mode spectra are measured in pure D$_2$O, for which the IR absorption in this spectral range is significantly smaller, we conclude that taking into account IR absorption is preferred, but failing to do so will impose an additional error of ~ 10 %, which might be acceptable depending on the aim of a study.

Figure 2C shows the $|\Gamma^{(2)}|^2$ input function for the surface response in the O-D stretch region as the blue dashed line. Using this function together with the IR pulse in Fig. 2A (red dashed line) and the 1 µm IR absorption spectrum of Fig. 1C, we compute the red line as $I_{SF,OD}$ using Eq. (1). It can be seen that this simulated spectrum differs significantly from the input: The main peak is shifted to lower frequencies, the spectral response is no longer symmetric, and at the high frequency side a distinct peak emerges. Processing this simulated equivalent of raw data by dividing it with the incident IR spectrum ($I_0(\omega_{IR})$) we obtain the blue curve. This processed spectrum has a completely different shape from the actual surface response ($|\Gamma^{(2)}|^2$, blue dashed line): there is a red-shifted peak (from 2500 cm$^{-1}$ to 2285 cm$^{-1}$), an overall asymmetric peaks shape, and there is a new peak-like feature at ~2710 cm$^{-1}$.

As water is known to have a significant non-resonant background[14] we also examine the effect of including this in $|\Gamma^{(2)}|^2$. We therefore incorporate a non-resonant contribution that has a magnitude of 1 % of the maximum response (Fig. 2D, blue dotted line). Taking this as the surface response, $I_{SF,OD}$ is similar to the one in Fig. 2C but with a more pronounced high frequency peak. Dividing this spectrum by the incident IR radiation ($I_{SF,CH}/I_0(\omega_{IR})$) we obtain the blue curve which is similar in the low frequency side to the one in Fig. 2C but now has a more distinct broad peak riding on top of a rising slope (highlighted in the inset).

Thus, linear absorption of IR infrared excitation pulses will drastically modify the measured SFS spectrum, such that it no longer agrees with the actual surface vibrational SF response. The ~ 160 cm$^{-1}$ blue shift of the main peak would lead one to conclude that the interfacial structure of water is more ordered or even ice-like than it actually is. Besides changes to the hydrogen bond network there is also the appearance of the high frequency peak at ~2700 cm$^{-1}$. This peak was not included in the original $|\Gamma^{(2)}|^2$ spectrum, and is therefore an artifact produced by IR absorption. Thus, without using the full form of Eq. (1) to correct SFS spectra from absorptive media, one would misinterpret the surface structure. Had the result of the present simulation been measured in a real experiment, and without applying Eq. (1) in reverse, one might have concluded that the interfacial hydrogen bond network is much more strongly connected than it is in reality and that the interface would be populated by dangling O-D groups.

**The effect of the SF light scattering volume.** Next, we examine the effect of the scattering volume, determined by the collection lens and the turbid medium (Fig. 1A). This lens has a light collection volume, which is determined by the specified focal length of the lens itself together with the linear optical scattering properties of the dispersion. In Eq. (1), this combined effect is described by $f_{focal}(z)$, which quantifies the ability to collect SF light from different spots. The number of SF photons that are collected from a specific $z$-positon depends on how far this spot is located away from the geometrical focal point of the lens, and what the effective focal distance of the lens is. Thus, the $z$-dependency of collection is convoluted with other depth-dependencies that are present in Eq (1). For a non-absorbing medium, the $z$-dependence of the collection optics does not play a big role since all scattering locations are illuminated with the same spectral profile. As a consequence, accounting for different collection efficiencies from different $z$-positions in the sample results in multiplying Eq. (1) by a constant. Spectral shapes or frequency distributions are not changed as a result. Therefore, thus far, SFS experiments from our laboratory have been performed with lenses that collect light from the entire optical path length of the sample (usually 100 – 1000 µm long, Refs.[45, 63]). This strategy ensures that as many SF photons as possible are collected onto the detector.

However, given that for an absorptive medium the IR light will be attenuated after several microns (Fig. 1C), and that this attenuation will be frequency dependent means that the spectral shape of the illuminating and scattered light will become $z$-dependent. It is therefore insightful to investigate the dependence of the emitted SF spectrum on the $z$-dependent intensity collection function $f_{focal}(z)$. One strategy of performing an experiment could be, instead of collecting all the emitted SF light, to collect only the light emitted from a thin slab just after the entrance window, since at this location the IR spectrum is least distorted. Therefore, we investigate a few scenarios: (1) - A collection $z$-range comprised of a thin 1 µm wide slab, positioned at 2 different positions in the sample (illustrated as the dark blue line in fig. 3A), (2) - A collection range with a wider Gaussian shape having 3 different widths (illustrated by the dashed lines in Fig. 3A), and (3) The commonly employed scenario in which all scattered SF light is collected.

The first scenario is achieved by setting $f_{focal}(z) = 1$ for $5 < z < 6\,\mu m$ and 0 elsewhere, and by setting $f_{focal}(z) = 1$ for $9 < z < 10\,\mu m$ and 0 elsewhere, representing two different locations close and a little further away from the input window. For $|\Gamma^{(2)}|^2$ we assume the spectral response of Fig. 2A (the blue line). Figure 3B shows the signal ($I_{SF,OD}$) collected from the entire sample region (same as the red line in Fig. 2C, plotted here as the red line), a 1 µm thick $z$-slab placed 5 µm deep into the sample cell (black line), and a 1 µm thick $z$-slab placed 10 µm deep into the sample cell (blue line). Comparing the scenario where all scattered light is collected (red line) to the 1 µm thick $z$-slab placed 5 µm deep into the sample, a difference between the SF detected light is found, comprised of a spectral shift on the low frequency side and a different high frequency shoulder on the high frequency side. The difference in spectral shape becomes more significant if the 1 µm slab is placed 10 µm deep into the sample cell (Fig. 3B, blue line), resulting in an SF spectrum that seems to have 2 peaks, rather than one. The difference between the two 1 µm $z$-slabs demonstrates that a small change in experimental parameters such as positioning of the collection lens can lead to dramatic changes in the detected spectral shape, even though the same interfacial response is detected. Although one could potentially calibrate the experiment for detecting the SF light from a narrow slab, given the $z$-sensitivity this strategy exposes one to unintentional errors.

Another option is to collect the intensity emitted from a wider range of sample depths, such as the ones sketched in Fig. 3A with a depth of focus (full width at $1/e^2$ of the maximum) of 50, 100 or 200 µm focused in the center of the sample cell, equivalent to using $f_{focal}(z) = e^{\frac{-(z-z_0)^2}{\Delta z_{focal}^2}}$ in Eq. (1). Figure 3C shows the computed spectra. The intensity from each slab is now additionally multiplied by individual collection coefficients defined by the collection profile of each lens (Fig. 3A, dashed, dashed-dotted and dotted lines). It can be seen that the smaller the collection volume the more impacted the resulting spectrum is. For a narrow $z$-range placed in the middle of the sample cell ($\Delta z_{focal}$= 50 µm, black line), only the high and low frequency side of the SF spectrum are collected as the IR absorption in these frequency ranges is less than in the central spectral range. Broadening it, by increasing $\Delta z_{focal}$ to 200 µm (twice the width of the sample cell, blue line) results in a detected SF spectrum that is very close to the theoretically correct SF spectrum. The above analysis therefore shows that choosing the widest possible collection volume (or $z$-range) is the best strategy for retrieving the most reliable surface SF spectrum from objects dispersed in water.

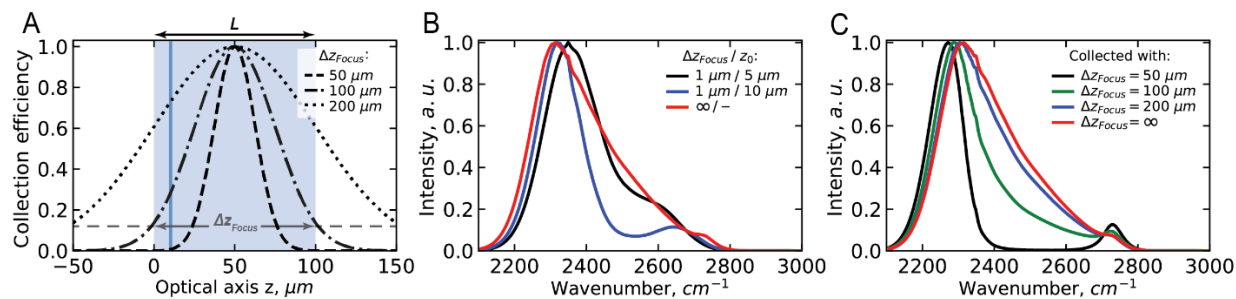

**Figure 3: SF light collection** (A) Sketch of the sample cell (light blue), with the profiles of the different $z$ −sectioning indicated: The vertical blue line represents a 1 µm thick slab, and the dashed, dot-dashed and dotted lines represent lenses with different collection depths. (B) Comparison of collective SF signal obtained from two slabs both 1 µm thick, but situated 5 µm deep (blue) and 10 µm deep (orange) from the entrance window of the sample cuvette. (C) Comparison of collective SF signal obtained from the whole 100 µm sample with different lenses (depth of focus is changed as indicated in panel A). The red line is simulated assuming there is no difference in the collection efficiency across the sample (i.e. $f_{focal}(z) = 1$), or in other words representing a lens with an infinite depth of focus.

**Conclusions**

We have analyzed the effect of linear optical processes on the outcome of sum frequency scattering measurements of particles dispersed in adsorptive medium. In particular, for the case of water, we found that infrared absorption drastically modifies the interfacial (O-D) vibrational stretch spectral response. We used a model spectral response for the second-order particle susceptibility that resembles a common single wide spectral distribution for the O-D stretch modes (similar to the Raman spectrum). The emitted SF intensity is computed taking into account IR absorption. The result of this computation, the simulated raw data of an SFS experiment is different from the original input spectrum: The main O-D stretch peak becomes asymmetric, and is red shifted significantly (by ~160 cm$^{-1}$) and a new peak-like feature appears in the high frequency range of the spectrum. This peak-like feature is found to be more pronounced in the presence of a non-resonant background. We also found that these spectral distortions cannot be avoided by varying the light collection optics. Selectively collecting the spectrum from a thin slab close to the sample cell entrance window or using a lens with smaller collection volume can both result in distorted spectra with red-shifted peak frequencies, as well as in peak-like feature at high and low spectral regions.

The simulations presented here illustrate that sum frequency scattering measurements performed in any absorptive medium, and in particular water, are incomplete without appropriate correction procedures for linear optical effects. The approach used here to simulate the effect of IR absorption and changes in the scattering volume can also be used to correct actual data, provided that the scattering volume of the collecting lens is measured, as well as the infrared spectrum of the bulk medium. The presented computation provides the framework for correction routines that should be implemented in future vibrational SFS experiments involving absorptive bulk media.

**Acknowledgements**

We thank Dr. Jerry I. Dadap for fruitful discussions. We thank the Julia Jacobi Foundation, and the Swiss National Science Foundation (grant numbers 200021-182606-1).